\documentclass[twocolumn,superscriptaddress,preprintnumbers,amsmath,amssymb,showpacs,floatfix]{revtex4-1}
\usepackage{graphicx}
\usepackage[export]{adjustbox}
\usepackage{dcolumn}
\usepackage{amsmath,amsbsy,amssymb,scalefnt}
\usepackage{bbm}
\usepackage{bm}
\usepackage{epstopdf}
\usepackage[T1]{fontenc}
\usepackage{textcomp}
\usepackage{mathtools}
\usepackage{breakcites}
\usepackage{hyperref}
\usepackage[mathscr]{eucal}
\usepackage{color,soul}
\soulregister\cite7
\soulregister\ref7
\usepackage[title]{appendix}
\usepackage[normalem]{ulem}
\newcommand\rst{\bgroup\markoverwith
{\textcolor{red}{\rule[.5ex]{2pt}{1pt}}}\ULon}
\setstcolor{red}

\DeclareMathAlphabet{\mathpzc}{OT1}{pzc}{m}{it}

\newcommand*{\Scale}[2][4]{\scalebox{#1}{$#2$}}%

\hypersetup{backref=true,
 pdfnewwindow=true, colorlinks=true,
 linkcolor=blue, anchorcolor=blue,
 citecolor=blue, filecolor=blue,
 menucolor=blue, urlcolor=blue}

\usepackage[below]{placeins}
\usepackage[usenames, dvipsnames, table]{xcolor}
\usepackage{siunitx}
\usepackage{float}
\usepackage{tabularx}

\usepackage{multirow}
\usepackage{booktabs}
\usepackage[utf8]{inputenc}
\usepackage{lmodern} 
\usepackage{diagbox}
\usepackage{makecell}
\usepackage{tikz}
\graphicspath{{./images/}}

\colorlet{khakestari.kam}{gray!30}
\colorlet{zard.mot}{yellow!70}
\colorlet{banafsh.mot}{violet!70}
\colorlet{banafsh.kam}{red!75!green!50!blue!25}
\colorlet{banafsh.kamtar}{red!75!green!50!blue!5}

\usepackage{colortbl}
\makeatletter

    \def\CT@@do@color{%
      \global\let\CT@do@color\relax
            \@tempdima\wd\z@
            \advance\@tempdima\@tempdimb
            \advance\@tempdima\@tempdimc
    \advance\@tempdimb\tabcolsep
    \advance\@tempdimc\tabcolsep
    \advance\@tempdima2\tabcolsep
            \kern-\@tempdimb
            \leaders\vrule
                    \hskip\@tempdima\@plus  1fill
            \kern-\@tempdimc
            \hskip-\wd\z@ \@plus -1fill }
    \makeatother

\begin{document}
\title{\bf The overlooked role of band-gap parameter in characterization of Landau levels in a gapped phase semi-Dirac system: the monolayer phosphorene case}
\date{\today} 
\author{Esmaeil Taghizadeh Sisakht}\thanks{taghizadeh.sisakht@gmail.com}
\affiliation{Department of Physics, Isfahan University of Technology, Isfahan, 84156-83111, Iran}
\author{Farhad Fazileh}
\affiliation{Department of Physics, Isfahan University of Technology, Isfahan, 84156-83111, Iran}
\author{S. Javad Hashemifar}
\affiliation{Department of Physics, Isfahan University of Technology, Isfahan, 84156-83111, Iran}
\author{Francois M. Peeters}
\affiliation{Department of Physics, University of Antwerp, Groenenborgerlaan 171, B-2020 Antwerpen, Belgium}


\begin{abstract}
Two-dimensional gapped semi-Dirac (GSD) materials are systems with a finite band gap that their charge carriers behave relativistically in one direction and Schrödinger-like in the other. In the present work, we show that besides the two well-known energy bands features (curvature and chirality), the band-gap parameter also play a crucial role in the index- and magnetic field-dependence of the Landau levels (LLs) in a GSD system. We take the monolayer phosphorene as a GSD representative example to explicitly provide physical insights into the role of this parameter in determining the index- and magnetic field-dependence of LLs.
We derive an effective one-dimensional Schrödinger equation for 
charge carriers in the presence of a perpendicular magnetic field and argue that the form of its effective potential is clearly sensitive to a dimensionless band-gap that is tunable by structural parameters. The theoretical magnitude of this effective gap and its interplay with oval shape $k$-space cyclotron orbits resolve the seeming contradiction in determining the type of the quantum Hall effect in the pristine monolayer phosphorene. Our results strongly confirm that the dependence of LLs on the magnetic field in this GSD material is as conventional two-dimensional semiconductor electron gases up to a very high field regime. Using the strain-induced gap modification scheme, we show the field dependence of the LLs continuously evolves into $B^{2/3}$ behavior, which holds for a gapless semi-Dirac system. The highlighted role of the band-gap parameter may affect the consequences of the band anisotropy in the physical properties of a GSD material, including magnetotransport, optical conductivity, dielectric function, and thermoelectric performance.
\end{abstract}


\maketitle
\footnotetext{\textit{$^{a}$~Address, Address, Town, Country. Fax: XX XXXX XXXX; Tel: XX XXXX XXXX; E-mail: xxxx@aaa.bbb.ccc}}
\section{Introduction}\label{intro5}
In the broad research area of condensed matter physics, the
theoretical and experimental study of the
integer quantum Hall effect (IQHE) and the fractional quantum Hall effect in exotic  
two dimensional (2D) materials are subjects of interest for many years.
The intense interest in this field may be attributed to the fact that
it can pave the way for more elucidation of many important features of 
quantum physics and interacting systems~\cite{novoselov2007room}.
In the IQHE regime and near the absolute zero, a 2D sample of high mobility electron gas
starts to deviate from the Drude model in both the Hall and the longitudinal resistance. The physics behind this phenomenon 
 is governed by the one-electron picture~\cite{laughlin1981quantized} and is in close relation with
 the formation of Landau levels (LLs) in the energy spectrum of the system~\cite{laughlin1981quantized}.
 In the one-electron scheme, it is well known that the curvature of energy bands and the chirality of charge carriers are two characteristics of the zero-field electronic spectrum that
 affect dramatically the index- and magnetic field-dependence of LLs, and thus the corresponding QHEs~\cite{zhang2005experimental,novoselov2006unconventional,dietl2008new,banerjee2009tight}.
These features lead to at least four known distinct classes of IQHE:
The first class describes the conventional QHE which is the characteristic of a traditional 2D 
electron gas or a 2D semiconductor system with a parabolic energy spectrum.
The second class  is the QHE in monolayer graphene that shows a quite
different unconventional character and is a manifest demonstration of the 
relativistic nature of charge carriers 
and their chirality (Berry’s phase~$\pi$). These factors leads to the LLs spectrum
of monolayer graphene
as $E_{n,\pm}=\pm\hbar\omega_c\sqrt{n}~(n=0,1,2,\cdots)$ where $\omega_c$ is the cyclotron
frequency and $n$ is the LL index.
Here, the important difference compared to the case of conventional QHE is
the existence  of a zero energy mode and
the unconventional form of the  
quantized Hall conductivity $\sigma_{xy}=\pm(n+\frac{1}{2})4e^2\texttt{\small/}h$,
which is a set of half-integer 
plateaus~\cite{novoselov2005two, zheng2002hall,gusynin2005unconventional}.
The third class is the unconventional QHE in multilayer graphene
where the low-energy spectrum of a
Bernal stacked j-layer graphene behaves as  $E(k)= k^j$ near the $\rm K$ and $\rm K'$ points.
When a low-energy charge carrier encircles a closed contour in the reciprocal space,
it gains a Berry's phase of $j\pi$.
By applying a perpendicular magnetic field to a  sheet of $j$-layer graphene,
its LLs are given by $E_{n,\pm}=\pm\hbar\omega_c\sqrt{n(n-1)(n-2)...(n-j)}~(n=0,1,2,\cdots)$~\cite{ezawa2013quantum}.
This gives rise to an extra $j$-fold degeneracy~\cite{novoselov2006unconventional}
in the zero-energy level (for  $E_0=E_1=...=E_j$), though for all other levels 
the degeneracy is 4-fold as in the monolayer case.
The fourth class refers to the QHE in the so-called gapless or 
zero-gap semi-Dirac (ZGSD) systems
that have a linear-quadratic electronic spectrum. It was shown that for such dispersions the dependence of LLs on the magnetic field is neither linear in the conventional
limit nor as $(nB)^{1\texttt{\small/}2}$ in the Dirac limit~\cite{dietl2008new,banerjee2009tight}. For these systems,
the LLs depend on the field as $[(n+1\texttt{\small/}2)B]^{2\texttt{\small/}3}$ and the absence of zero energy mode is due to the 
cancellation of the Berry's phase~\cite{dietl2008new,banerjee2009tight}.

2D semi-Dirac materials can be divided into two distinct 
general types of zero-gap and gapped phases. 
A  zero-gap semi-Dirac (ZGSD)  system has a semi-metallic character whose charge carriers behave as massless Dirac fermions along one direction and are Schrödinger-like along the other. In comparison, for a gapped semi-Dirac (GSD) system, the dispersion of charge carriers is relativistic along one direction and parabolic along the other. 
In ZGSD systems, the dominant reason for exhibithing novel features is thier highly anisotropic band dispersion,
in the absence of any band-gap. Several systems are proposed to host ZGSD fermions. Some examples are honeycomb lattice models~\cite{dietl2008new,banerjee2009tight,montambaux2009merging}, TiO$_2$/VO$_2$ heterostructures~\cite{pardo2009half}, organic conductor $\alpha$-(BEDT-TTF)$_2$I$_3$ salt~\cite{katayama2006pressure}, strain- or electric field-induced semimetal phase in few-layer black phosphorus~\cite{baik2015emergence,liu2015switching,wang2015strain}, and experimentally realization in potassium doped few-layer phosphorene~\cite{kim2015observation}.
Some characteristic features of such materials include the mentioned unusual LLs and QHE, highly anisotropic longitudinal optical conductivity~\cite{carbotte2019optical}, highly anisotropic plasmon frequency~\cite{banerjee2012phenomenology}, and the power-law decay index $-5\texttt{\small/}4$ for local density of states oscillations~\cite{chen2021power}. On the other hand, for GSD materials, although the band anisotropy is reflected in many properties, it is not the only decisive factor. In other words, there are characteristic features in such systems that cannot be explained only by using the anisotropy in dispersion of electronic bands.
For example, introducing a band-gap in a honeycomb semi-Dirac lattice leads to a completely different optical conductivity~\cite{carbotte2019optical}. It has also been shown that the power-law decay index for local density of states oscillations in a GSD system 
is $-1\texttt{\small/}4$ and identified as a fingerprint of the system~\cite{chen2021power}.
These results show that the band-gap parameter also comes into play in such systems. Therefore, a reliable description of the interplay of the band-gap parameter with the highly anisotropic band spectrum to highlight its role in determining characteristic features is of key importance.

Based on our general classification, single- and few-layer 
phosphorene are GSD materials
that their QHEs and LLs have come into focus  in recent  years~\cite{li2015quantum,li2016quantum,long2016achieving,tran2017surface,
yang2017integer,zhou2015landau,pereira2015landau,yuan2016quantum,lado2016landau,ezawa2015highly,jiang2015magnetoelectronic}.
 The highly anisotropic physics in many properties of phosphorene are closely related to
 its GSD band structure~\cite{rodin2014strain,low2014tunable,ezawa2014topological,ezawa2015highly}. 
 By applying a finite magnetic field B, oval shape (ellipse-like) $k$-space cyclotron orbits form in this 2D material~\cite{ezawa2015highly} and 
 one might expect to observe an anomalous behavior as $B^{2\texttt{\small/}3}$ in the LLs field dependence. 
 But, there exist other theoretical and experimental works~\cite{zhou2015landau,pereira2015landau,yuan2016quantum,lado2016landau,jiang2015magnetoelectronic,yang2017integer}
 showing a linear field dependence of LLs in phosphorene, similar to the conventional isotropic semiconductor electron gases.
 Therefore, considering the mentioned highly anisotropic dispersion and consequently the formation of ellipse-like cyclotron orbits, a seeming contradiction appears in the field dependence of LLs on the magnetic field.
  In the present paper, we resolve this contradiction  by deriving
  an effective one dimensional Schr\"{o}dinger equation for 
  charge carriers of monolayer phosphorene (MLP) in the 
  presence of a perpendicular magnetic field and clarify the underlying physical mechanism governing the magnetic field dependence of LLs.
  To derive the mentioned equation, we employ simple yet accurate tight-binding (TB) 
  and continuum models near the Fermi energy of MLP. 
  Our  approach highlight in a transparent way the 
  competitive role of an effective dimensionless gap (originated from the band-gap parameter) and the curvature of the electronic bands in determining this characteristic feature.
  Our studies reveal that
   the theoretical magnitude of this effective gap is a decisive factor 
   that makes the behavior of LLs in MLP  as conventional 2D semiconductor electron gases 
   at least up to a very high field regime.
   Next, we will discuss the conditions under which such dependence can
   continuously evolve to the case of a ZGSD system. Our results not only provide 
   a simple vision for understanding the formation of  LLs in MLP, but also 
   unveil that the band-gap parameter 
   plays an important (hidden) role in mediating
   the type of IQHEs in 2D semi-Dirac materials. This is taken as an indication that the band-gap parameter might manifests
   itself in many other properties.
\section{ Effective low-energy Hamiltonian of MLP}
\begin{figure}[!t]
\centering
{\includegraphics[angle=0,width=0.28\textwidth]{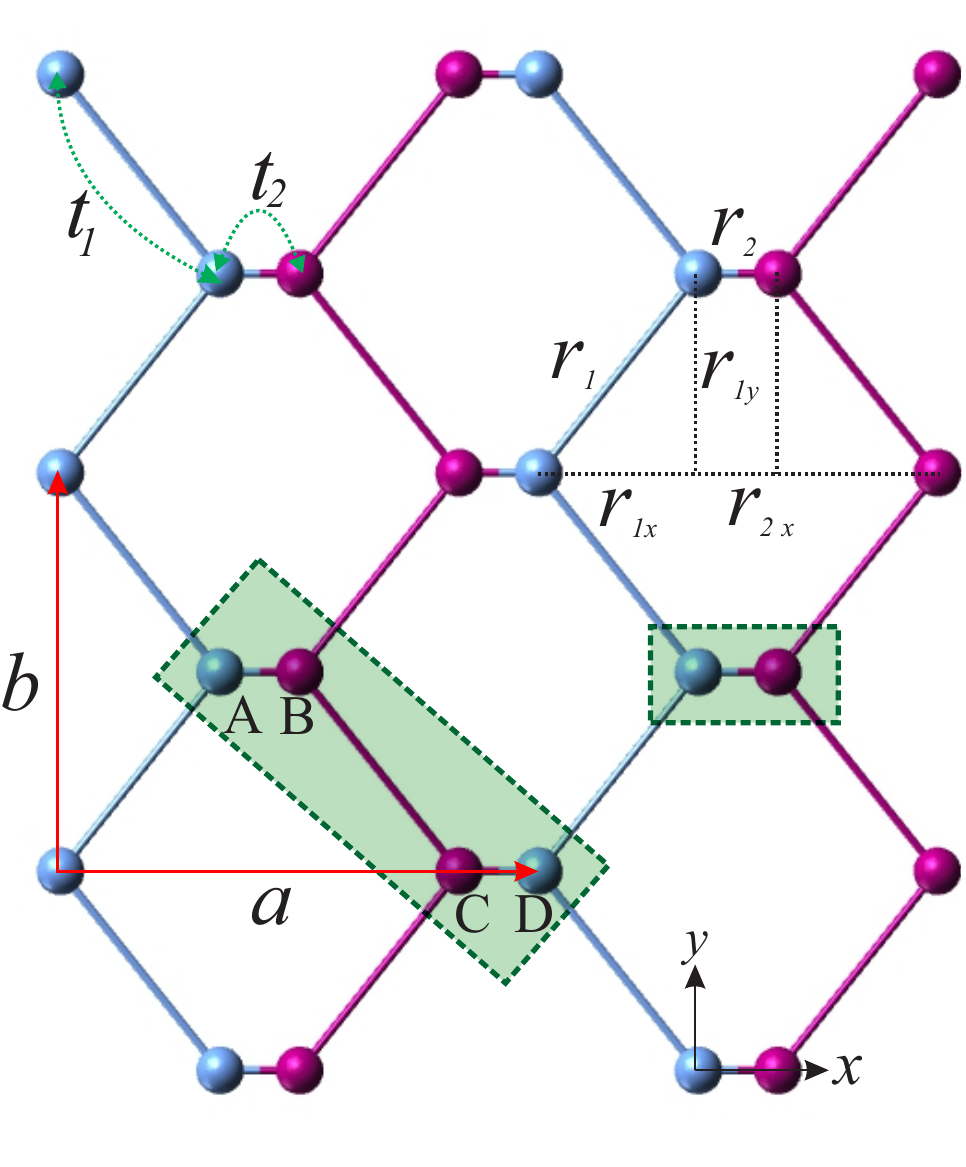}}
 \caption
 { Lattice geometry of monolayer phosphorene. Two different colors of P 
 atoms indicate the upper and lower chains of MLP. Lattice vectors, components of 
 geometrical parameters, and dominant hopping parameters $t_1$ and $t_2$ are shown.  
 Green dashed rectangle shows the unit cell of MPL including four P atoms mark as A,B, C and D. 
 }
\label{lattice}
\end{figure}

\begin{figure}[!t]
\centering
{\includegraphics[angle=0,width=0.4\textwidth]{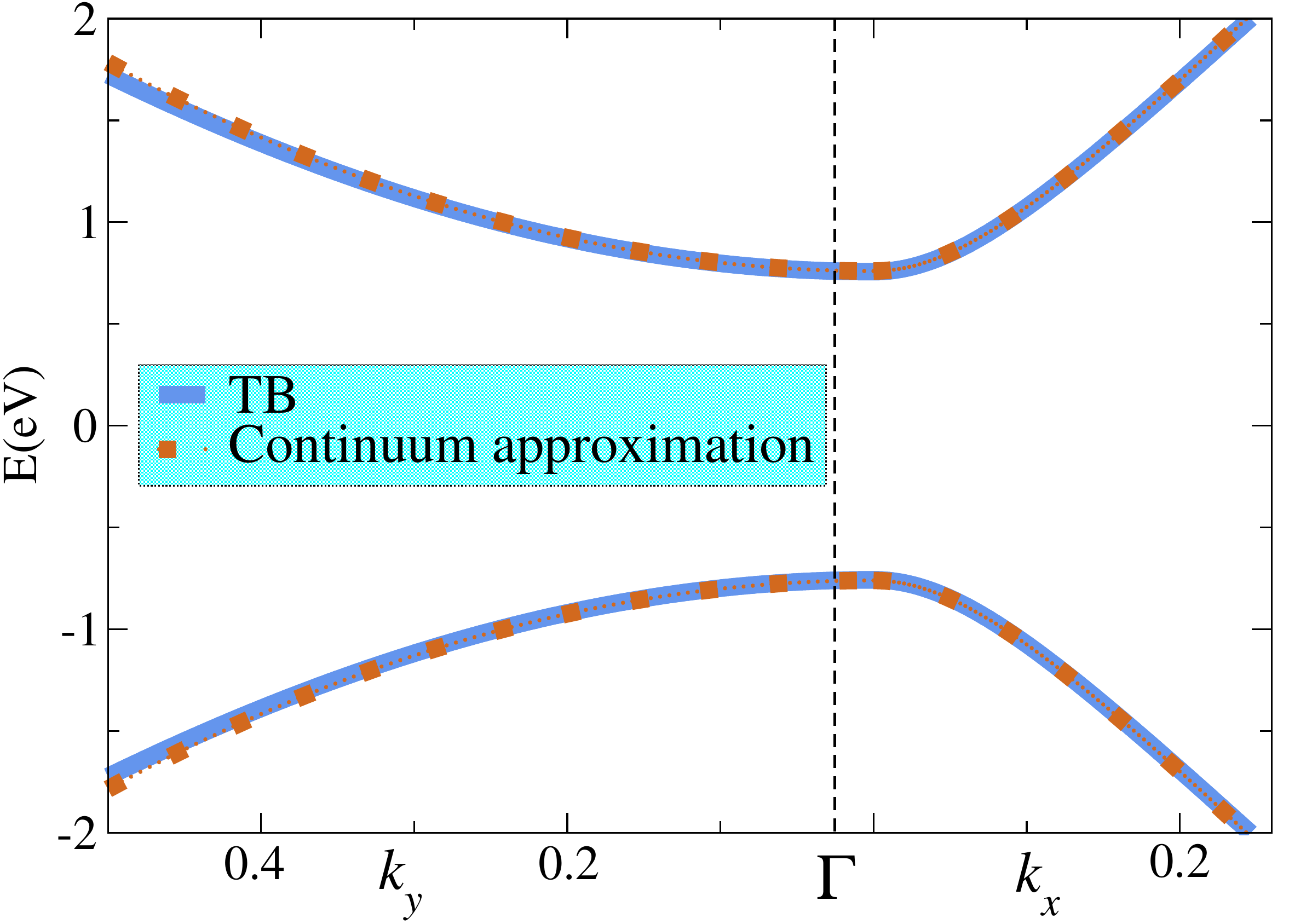}}
 \caption{ Low-energy spectrum of MLP obtained from the 
 continuum approximation and the TB model.}
\label{low}
\end{figure}

Figure.~\ref{lattice}  shows the lattice geometry and structural parameters of 
MLP.  We have chosen the $x$
and $y$ axes along the armchair and zigzag directions, respectively and the $z$ 
axis in the normal direction to the plane of phosphorene.
With this definition of coordinates, one can indicate the various atom connections $r_i$.
The structure parameters have been taken from~\cite{qiao2014high} that are
very close to the experimentally measured parameters~\cite{takao1981electronic} of the bulk structure.
The components of the geometrical parameters for bond lengths 
$r_1=2.240$ \r{A} and $r_2=2.280$ \r{A}
are $(r_{1x},r_{1y},r_{1z})=(1.503,1.660,0)$ and $(r_{2x},r_{2y},r_{2z})=(0.786,0,2.140)$, respectively.
The two in-plane lattice constants are $a=4.580$ \r{A}, and $b=3.320$ \r{A}, 
and the thickness of a single layer due to the puckered nature is $r_{2z}=2.140$ \r{A}.
There exist many studies~\cite{rudenko2014quasiparticle,rudenko2015toward,popovic2015electronic}
showing that the low-energy electron states near the gap region are well described by considering 
only one effective $p_z$-like orbital per lattice 
site.  
The proposed single-orbital tight-binding (TB) model for this GSD system in Ref.~\cite{rudenko2015toward} involves ten 
different hopping parameters that are also usable as intra-layer hopping parameters of the 
few-layer phosphorene or its bulk structure. However, it is  
shown~\cite{rudenko2014quasiparticle,rudenko2015toward,ezawa2014topological,popovic2015electronic}
that the main aspects of the low-energy spectrum of this GSD material, such as energy gap 
evaluation and bands curvature are
well described by only two dominant energy hopping integrals  $t_1$ and $t_2$ (see Fig.~\ref{lattice}).
These parameters are defined as the hopping between the nearest neighbors along the zigzag and 
armchair directions, respectively. Moreover, the curvature of energy bands is a salient 
factor that also plays role in the field dependence of LLs in a 
2D semi-Dirac system~\cite{dietl2008new}. Therefore, it is good enough for our purpose to 
suppose the TB Hamiltonian of MLP as
\begin{equation}
\hat{H}=\sum_{ i,j }t_{ij}c_{i}^{\dagger }c_{j},
\label{H15}
\end{equation}
where $c_i^{\dagger}$ and $c_j$ are the creation and annihilation operators of $p_z$-like orbitals at 
sites $i$th and $j$th, respectively,  and the hopping parameters $t_{ij}$ run only 
over the two hopping parameters $t_1$ and $t_2$. As we shall see in the
following, such an approximation simplifies our analyses in addition to the 
fact that it does not affect our results based on the above mentioned reasons.
The unit cell of MLP is a rectangle containing four phosphorus (P) atoms as shown in 
Fig.~\ref{lattice} and labeled by A, B, C, and D. Fourier transform of the Eq.~(\ref{H15}) gives the 
general Hamiltonian in momentum space as
\begin{equation}
H=\sum_{\bm k} \psi^{\dagger}_{\bm k}H(\bm k)\psi_{\bm k},
\label{H25}
\end{equation}
where we have used the basis  
$\psi^{\dagger}_{\bm k}=\{A^{\dagger}_{\bm k},B^{\dagger}_{\bm k},C^{\dagger}_{\bm k},D^{\dagger}_{\bm k}\}$
with $A^{\dagger}_{\bm k}=\mathcal{N}^{-1\texttt{\small/}2}\sum_{i} e^{i\bm k\cdot\bm {r_i}}A^\dagger_i$, and so on. 
Here, $\bm k$ is the Bloch momentum, and $\bm {r_i}$ is the orbital position with respect to the origin of 
coordination system located at the position of atom A. The kernel Hamiltonian $H(\bm k)$ in Eq.~(\ref{H25})
is represented as
\begin{eqnarray}
H(\bm k) &=&\begin{pmatrix} 0 & h_{12}(\bm k) & 0 & h_{14}(\bm k)  \\
h^*_{12}(\bm k)& 0 & h^*_{14}(\bm k)  & 0  \\ 
0 & h_{14}(\bm k) & 0 & h_{12}(\bm k)  \\
h^*_{14}(\bm k) &0 & h^*_{12}(\bm k) & 0  
\end{pmatrix},
\label{H35}
\end{eqnarray}
whose elements are given by
\begin{eqnarray}
 h_{12}(\bm k) &=& t_2e^{ik_xr_{2x}}\notag \\
 h_{14}(\bm k) &=& 2t_1e^{-ik_xr_{2x}}\cos(k_yr_{1y}).
 \label{H411}
\end{eqnarray}
\begin{figure}[!t]
\centering
{\includegraphics[angle=0,width=0.4\textwidth]{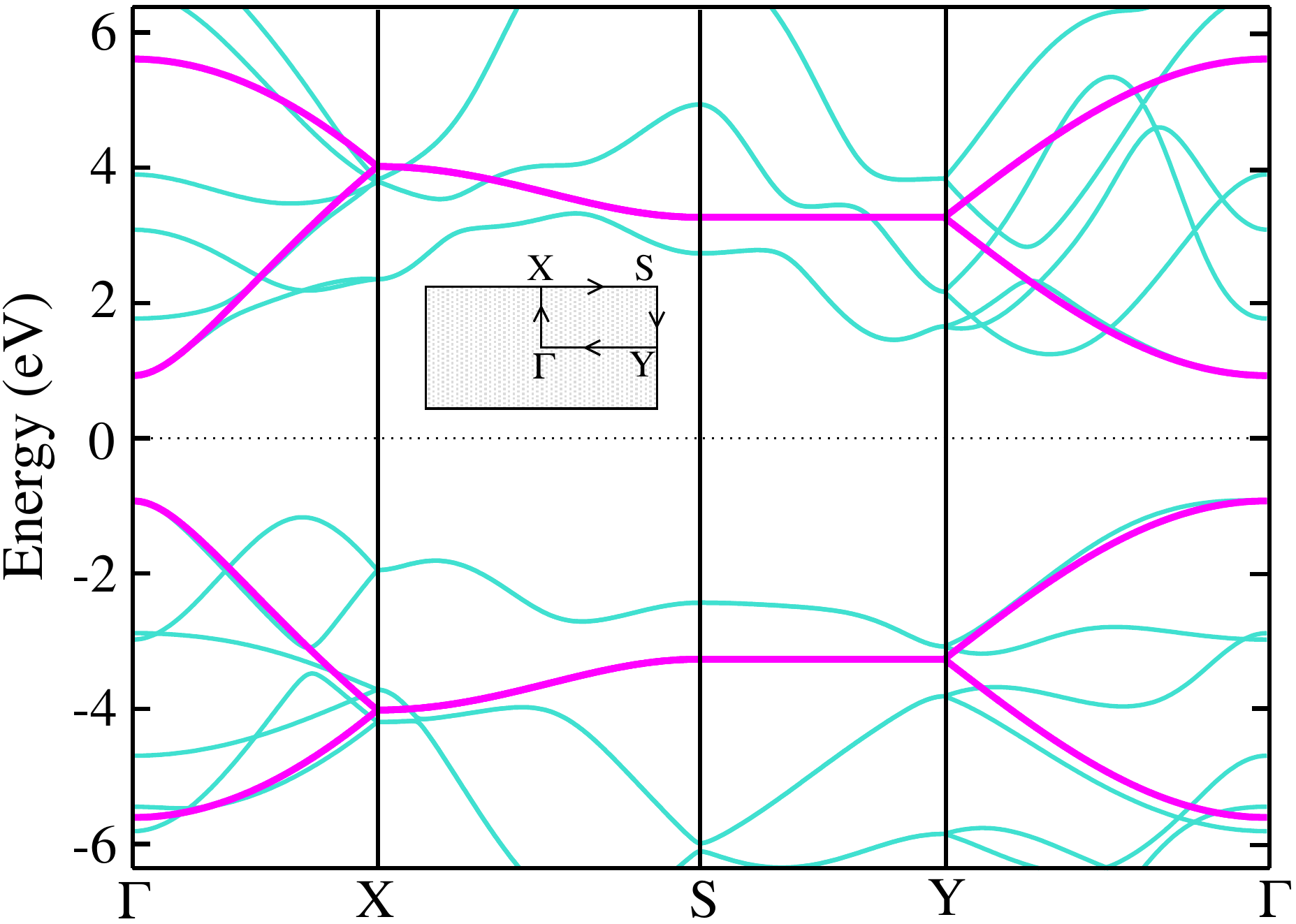}}
 \caption{
 Comparison of low-energy bands of MLP along
symmetry directions  as obtained from the TB Hamiltonian (magenta lines)
and the band structures as calculated by DFT (cyan curves) taken from~\cite{rudenko2015toward}.
}
\label{fit}
\end{figure}
Using the Eqs.~(\ref{H35}) and (\ref{H411}), we have obtained the two hopping
integrals $t_1$ and $t_2$ by simply fitting the TB bands of phosphorene with the 
DFT data~\cite{rudenko2015toward} (see Fig.~\ref{fit}). The obtained numerical values of these
parameters are $t_1=-1.170$ eV and $t_2=3.267$ eV.
As shown, this four-band model describes very well the curvature of
the valance and conduction bands near the gap of the system.
This  demonstrates the validity of the model near the gap region,
allowing us to remarkably simplify our calculations for this typical GSD material.
Due to the nonsymmorphic $D_{2h}$ point group invariance of MLP
lattice~\cite{kurpas2016spin}, one can reduce the four-band model   
to a two-band Hamiltonian as
\begin{flalign}
 \mathcal H(\bm k)& = \begin{pmatrix} 0 & h_{12}(\bm k)+h_{14}(\bm k) \\
 h^*_{12}(\bm k)+h^*_{14}(\bm k)& 0 
\label{hii}
\end{pmatrix}.
\end{flalign}
By expanding the Hamiltonian matrix elements in the vicinity of $\Gamma$ point,
one can write a continuum approximation. Retaining the terms up to the 
leading non-zero order of $k_x$ and $k_y$, the continuum model is given by
\begin{equation}
\mathcal H(\bm k)=(\alpha+\beta k_{y}^2)\bm\sigma_x-\gamma k_x\bm\sigma_y,
\label{Hk21}
\end{equation}
demonstrating that the spectrum has relativistic nature along the $k_x$ direction, while it is parabolic
along the $k_y$ direction. Here, $\bm\sigma_x$ and $\bm\sigma_y$ are Pauli 
matrices and $\alpha$, $\beta$ and 
$\gamma$ are given by
\begin{eqnarray}
 \alpha &=& 2t_1+t_2 ,\notag \\
 \beta &=& -t_1r_{1y}^2, \notag \\
 \gamma &=& -2t_1r_{1x}+t_2r_{2x},
 \label{str}
\end{eqnarray}
whose numerical values are $\alpha=0.925$ eV, $\beta=3.20$ eV$\cdot$\AA$^2$ and $\gamma=5.765$ eV$\cdot$\AA.
The continuum spectrum relations for electrons and holes are then simply given by
\begin{equation}
 E(k_x,k_y)=\pm\sqrt{(\alpha+\beta k_y^2)^2+\gamma^2k_x^2},
\label{Ek_xy}
\end{equation}
with an energy band gap of $E_g=2\alpha=1.85$ eV at the $\Gamma$ point.
A comparison between the low-energy spectrum of the TB and continuum approximation is shown
in Fig.~\ref{low}. There exists an excellent agreement between the two approaches for energies
in the range of $-1.5$ to $1.5$ eV. Note that, due to  ignoring the further 
hopping terms~\cite{rudenko2015toward},  we have lost  the 
weak  electron-hole asymmetry of the  energy spectrum. This reflects itself 
as a small error in  the effective masses of  electrons and holes~\cite{pereira2015landau}. 
Thus, from the dispersion relation of Eq.~(\ref{Ek_xy}) the effective masses of electrons and 
holes for the $x$ and $y$ directions are estimated as
\begin{equation}
 m_x^{e,h}=\pm\frac{\hbar^2}{2\gamma^2},~~~~~m_y^{e,h}=\pm\frac{\hbar^2}{2\beta},
\label{mass}
\end{equation}
where the resulting effective masses in terms of the free electron mass $m_{\scriptscriptstyle 0}$ are 
$m_x^{e,h}=\pm0.11m_{\scriptscriptstyle 0}$ and $m_y^{e,h}=\pm1.19m_{\scriptscriptstyle 0}$, consistent with other 
results~\cite{pereira2015landau,sisakht2015scaling}. This implies that our approximation captures 
the existence of a strong
anisotropy in the low-energy electronic spectrum of this GSD system which is essential for properly predicting of main distinguishing features.
\section{The role of band-gap parameter  in Landau levels of MLP}\label{LLsphos}
\begin{figure}[!t]
\centering
{\includegraphics[angle=0,width=0.45\textwidth]{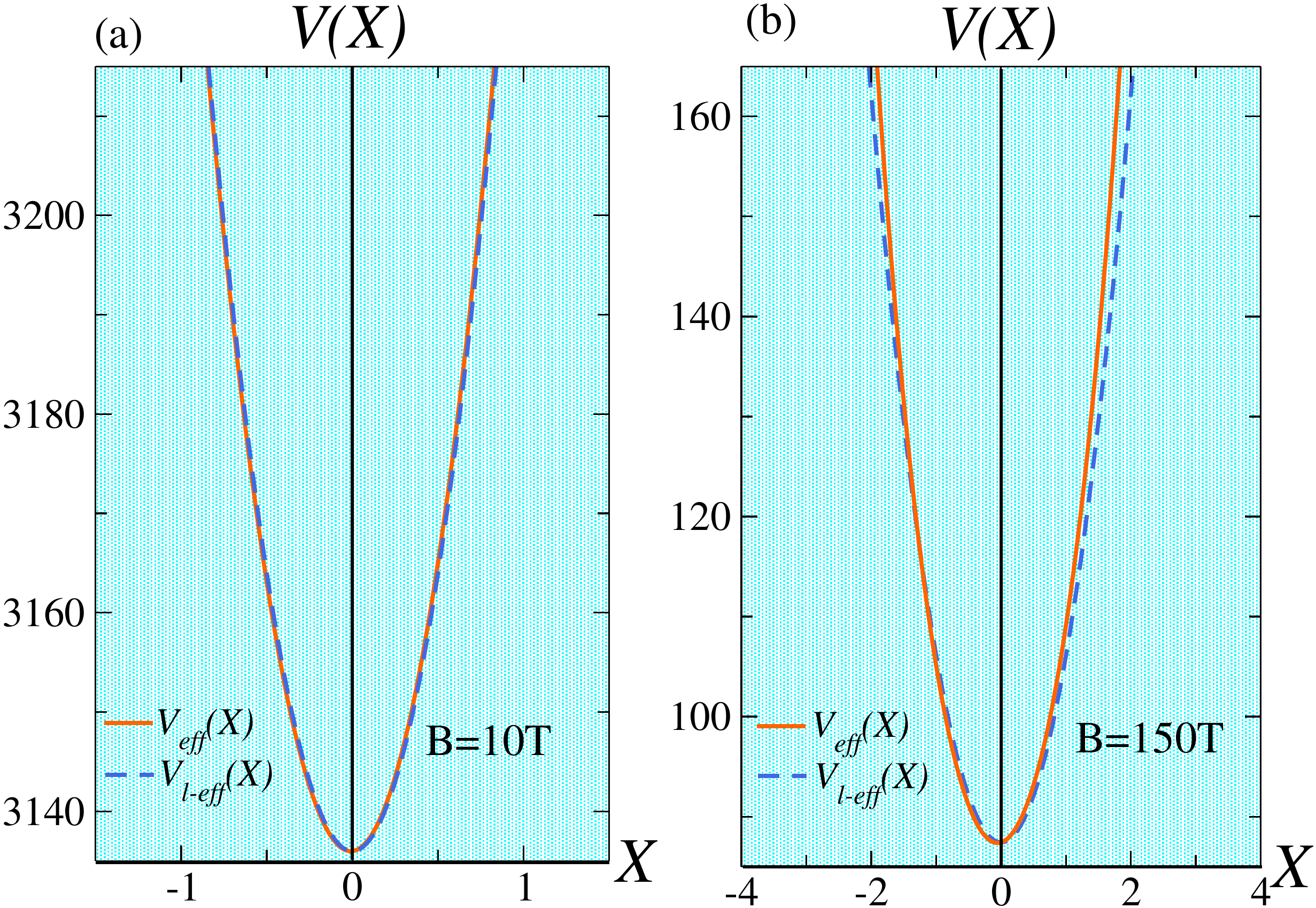}}
 \caption{ Comparison of the effective potentials $V_{eff}(X)$ and $V_{l\Scale[0.6]{-}eff}(X)$ at (a) $B$=10~T
 and (b) $B$=150~T. }
\label{potential}
\end{figure}

Charge carriers with dispersion Eq.~(\ref{Ek_xy}) 
form oval shape $k$-space orbits in the presence of a perpendicular uniform magnetic field $B$. Based on the Onsager quantization constraint
on the $k$-space area of a closed cyclotron $S(E)$, it is shown that the index- and field-dependence of LLs obey the power law $2\texttt{\small/}3$~\cite{ezawa2015highly}. Also, it is demonstrated~\cite{dietl2008new,banerjee2009tight} that such a power law holds for  ZGSD systems 
in which they have the same $k$-space cyclotron orbits as GSD materials. 
However, there are both theoretical and experimental evidences indicating that this field dependence for MLP is not correct~~\cite{zhou2015landau,pereira2015landau,yuan2016quantum,lado2016landau,jiang2015magnetoelectronic,yang2017integer} in a GSD system, and thus a seeming contradiction appears here. In what follows, we resolve this contradiction by deriving an effective gap-dependent one-dimensional Schrödinger equation for charge carriers of this GSD material in the presence of a perpendicular magnetic field. 
Let us consider the continuum Hamiltonian of  MLP in the presence of
magnetic field $\bm B$. Using the gauge $\bm{A}=(0,Bx,0)$
and the substitution $k_y \rightarrow k_y+\ell_B^{-2}x$, the new Hamiltonian is given by
\begin{equation}
\mathcal H(\bm k,B)=(\alpha+\beta \eta^2\tilde{x}^2)\bm{\sigma}_x-\gamma k_x\bm{\sigma}_y.
\label{HKB}
\end{equation}
Here, $\ell_B$ is the magnetic length which is defined as $\ell_B^{-2}\equiv\eta=eB/\hbar$, and
we have defined the new variable $\tilde{x}=k_y\texttt{\small/}\eta-x$.
Squaring the Eq.~(\ref{HKB}), and after some algebraic  calculations we arrive at
\begin{equation}
\mathcal H^2(\bm k,B)=\gamma^2 k_x^2\mathbbm{1}+(\alpha+\beta \eta^2\tilde{x}^2)^2\mathbbm{1}+i\beta \gamma\eta [k_x,\tilde{x}^2]\bm\sigma_z,
\label{HkB1}
\end{equation}
where $\sigma_z$ is the $z$-component of the Pauli matrices. Using the commutation 
relation $[k_x,\tilde{x}]=-i$, we define the 
dimensionless variables $X=\tilde{x}\texttt{\small/}\varepsilon$ and $P=\varepsilon k_x$ so that they satisfy
the commutation relation $[P,X]=-i$. Substituting these new variables in 
Eq.~(\ref{HkB1}) gives  
\begin{equation}
\mathcal H^2(\bm k,B)=C\big(P^2+(\delta+X^2)^2\mathbbm{1}+i[P,X^2]\bm\sigma_z\big),
\label{HKB2}
\end{equation}
where $C=(\beta\eta^2\gamma^2)^{2\texttt{\small/}3}$, $\delta=\alpha\texttt{\small/}\sqrt{C}$, and $[P,X^2]=-2iX$.
This implies that it is enough to solve an effective Schr\"{o}dinger equation with the effective
potential of
\begin{equation}
V_{eff}(X)=(\delta^2+2\delta X^2+X^4)\mathbbm{1}+2X\bm\sigma_z,
\label{v_eff1}
\end{equation}
where the dimensionless variable $\delta$ acts as an effective gap and equals to
\begin{equation}
\delta=\frac{\alpha}{[\beta\gamma^2B^2(e/\hbar)^2]^{1\texttt{\small/}3}}=\frac{E_g}{2[\beta\gamma^2B^2(e/\hbar)^2]^{1\texttt{\small/}3}}.
\label{delta1}
\end{equation}
Thus, the eigenvalues of the Schr\"{o}dinger equation 
$\mathcal H(\bm k,B)\Psi^{e,h}_n=\mathcal{E}_n\Psi^{e,h}_n$ are simply related to the 
eigenvalues $E_n$ of the square Hamiltonian~(\ref{HKB2}) via
\begin{equation}
\mathcal{E}_n=\pm(\beta\eta^2\gamma^2)^{1\texttt{\small/}3}\sqrt{E_n}.
\label{eps_n}
\end{equation}
The low-energy barrier 
shape of the effective potential~(\ref{v_eff1})  (with a quartic form)
is strongly dependent on the effective gap $\delta$, making it a crucial factor in determining the field dependence of LLs on the magnetic field.
This implies that in addition to the bands curvature, the energy band gap is also an
important factor in determining field dependence of LLs.
In order to calculate the LLs spectra of MLP, one can substitute the magnetic
length $\ell_B=256.5\texttt{\small/}\sqrt{ B}$~\AA ~(the magnitude of
$B$ is written in the unit of Tesla), and
the numerical values of the structural parameters~(\ref{str}) in Eq.~(\ref{delta1}) to obtain the
dimensionless gap $\delta$ as
\begin{equation} 
\delta\approx\frac{264.0~}{ B^{2\texttt{\small/}3}}.
\label{delta2}
\end{equation}
This shows that even in the presence of a whopping strength of magnetic field up to $\sim150$T we have
$\delta\gg1$.
Figures~\ref{potential} (a) and (b)  illustrate  the comparison between the $V_{eff}(X)$ and the low-energy limit of effective potential
\begin{equation}
V_{l\Scale[0.6]{-}eff}(X)=\delta^2+2\delta X^2,
\label{V_leff}
\end{equation}
at magnetic fields $B=10$~T and $B=150$~T, respectively. As clearly seen, for a high magnetic field $B=10$~T, they precisely coincide and for the ultra-high magnetic field $B=150$~T,
we found no significant deviation. 
Thus, even in the presence of a very strong magnetic field, it is sufficient to find
the energy levels of
\begin{equation}
\big(-\frac{d^2}{dX^2}+(\delta^2+2\delta X^2)\big)\psi_n=E_n\psi_n.
\label{En1}
\end{equation}
This equation has a quadratic form with the spectrum of
\begin{equation}
E_n=\delta^2+2\sqrt{2\delta}(n+\frac{1}{2}).
\label{En2}
\end{equation}
Substituting this spectrum in Eq.~(\ref{eps_n}) we arrive at
\begin{eqnarray}
 \mathcal{E}_n  &=& \pm\alpha\big(1+\frac{2\sqrt{2}}{\delta \sqrt{\delta}}(n+\frac{1}{2})\big)^{1/2} \notag \\
  &\approx& \pm\big(\alpha+\frac{\sqrt{2}}{\sqrt{\delta}}\frac{\alpha}{\delta}(n+\frac{1}{2})\big)\notag \\
  &=&\pm\big(\alpha+\omega_c(n+\frac{1}{2})\big),
 \label{En3}
\end{eqnarray}
where we have used Eq.~(\ref{mass}) to define  $\omega_c=eB\texttt{\small/}\sqrt{m_xm_y}$.
This result explicitly demonstrates that the LLs dependence on magnetic field 
 resembles the behavior of Schr\"{o}dinger electron gases. 
 As seen, the results are
valid up to the very high field regime.
\section{Continuous evolution of the field dependence of Landau levels}\label{LLsphos}
\begin{figure}[!t]
\centering
{\includegraphics[angle=0,width=.47\textwidth]{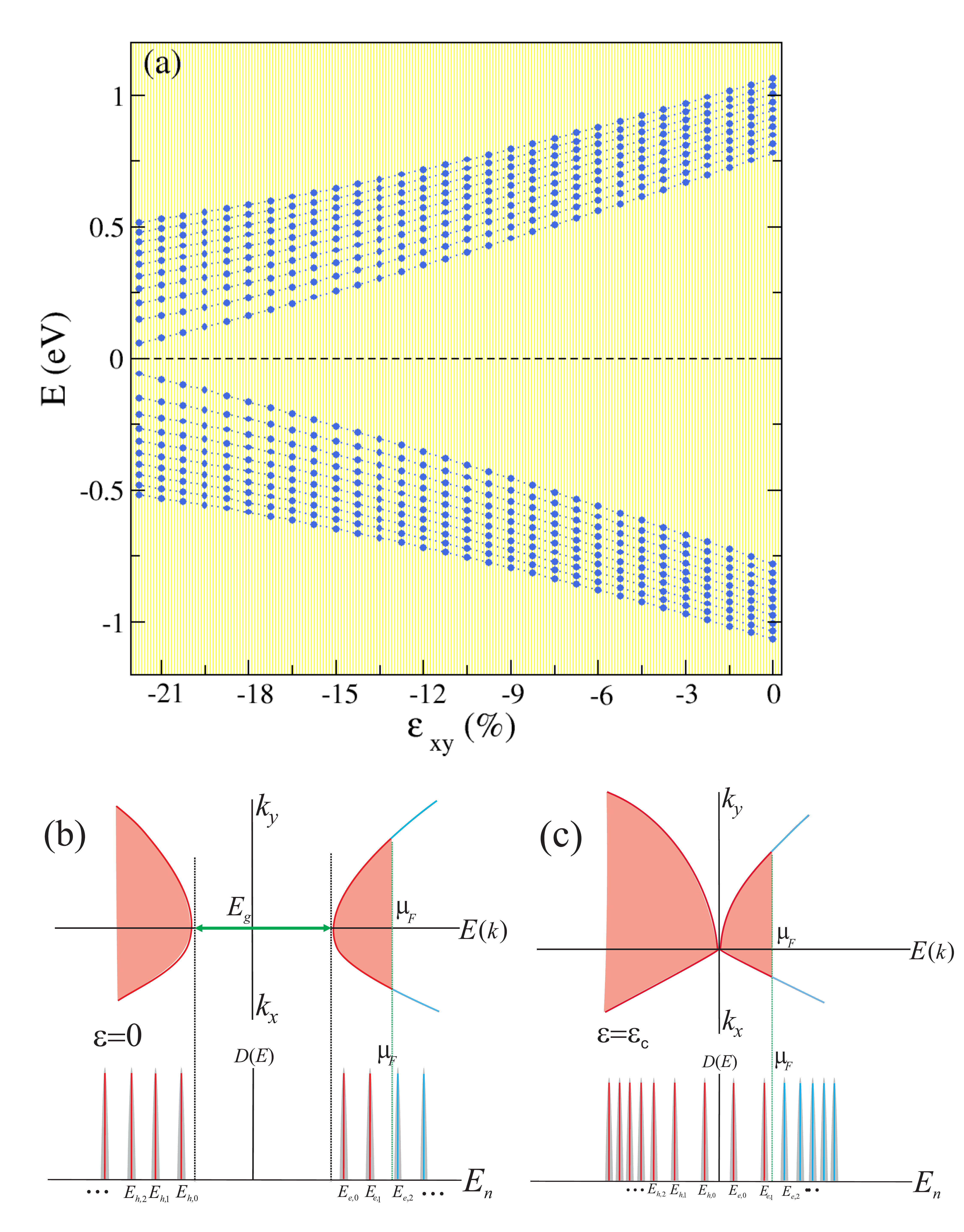}}
 \caption{ (a) LLs of MLP as 
a function of equi- biaxial strain. (b) and 
(c) show equidistant LLs for unstrained phosphorene and $B^{3\texttt{\small/}2}$ dependence of LLs at the critical value $\varepsilon_c$,
respectively.}
\label{ll_mono}
\end{figure}

The tunability of the band-gap parameter and hence the effective potential (\ref{v_eff1})  allows us to continuously evolve the field dependence of LLs in our GSD system. 
The role of uniaxial and biaxial strain in manipulating the electronic structure of single- and few-layer phosphorene been investigated
via DFT~\cite{rodin2014strain,peng,wang,zhang2015stacked,huang2014semiconductor} and 
TB approaches~\cite{jiang2015anal,mohammadi,duan}. Applying 
tensile or compressive strain 
in different directions results in different modifications of the electronic bands. One can observe a 
direct to indirect gap transition, or a prior direct 
band gap closing, depending on the type of the applied strain~\cite{peng,wang,zhang2015stacked}.
Here, we consider equi-biaxial  compressive strain in the plane of MLP.
This modifies the low energy bands so that
the valence and conduction bands approach each other.
Investigating the local density of states of $p$ orbitals~\cite{zhang2015stacked} shows that the used
one orbital $p_z$-like TB model is still valid in the low energy limit.
In addition, when an in-plane equi-biaxial strain ( $\varepsilon_x=\varepsilon_y=\varepsilon$ ) is applied to phosphorene, the rectangle shape of the unit cell remains intact, and therefore the nonsymmorphic $D_{2h}$ point group of the system does not change. 
Hence, our low-energy continuum model is successful to explain the manipulation of energy bands in the presence of an equi-biaxial compressive strain and provides a way to investigate the evolution of  LLs in MLP.
It is shown that both bond lengths and bond angles of MLP change under axial strains~\cite{wang,sa2014strain}.
According to the Harisson  rule~\cite{harrison2004elementary}, 
the hopping parameters for $p$ orbitals are  related to the bond length
as $t_i\propto 1\texttt{\small/}r_i^2$ and the angular dependence can be described by the hopping 
integrals along
the $\pi$ and $\sigma$  bonds.
However, our calculations showed that, although the changes in angles are almost 
noticeable~\cite{wang,sa2014strain}, their modification of the hopping parameters is much weaker than that of bond lengths. Hence, we only consider the changes of bond lengths in the hopping modulation.
In the presence of biaxial strain $\varepsilon$ the initial geometrical parameter $r^{\scriptscriptstyle 0}_i$ (corresponding to initial lattice constants $a_{\scriptscriptstyle 0}$ and $b_{\scriptscriptstyle 0}$)
is deformed 
as $(r_{ix}, r_{iy})=(1+\varepsilon)(r^{\scriptscriptstyle 0}_{ix},r^{\scriptscriptstyle 0}_{iy})$. In the linear 
deformation 
regime, expanding the
norm of $r_i$ to the first order of $\varepsilon$ gives
\begin{equation}
r_{i}=[1+(\alpha_x^i+\alpha_y^i)\varepsilon]r^{\scriptscriptstyle 0}_i,
\end{equation}
where $\alpha_x^i=(r^{\scriptscriptstyle 0}_{ix}/r^{\scriptscriptstyle 0}_i)^2$, 
and $\alpha_y^i=(r^{\scriptscriptstyle 0}_{iy}/r^{\scriptscriptstyle 0}_i)^2$.
Using the Harrison relation,
we obtain the  effect of strain on the hopping parameters as
\begin{equation}
t_i\approx[1-2(\alpha_x^i+\alpha_y^i)\varepsilon)]t^{\scriptscriptstyle 0}_i,
\label{MODIFIED_t0}
\end{equation}
where $t_i$ is a modified hopping parameter of deformed phosphorene with new lattice constants $a$ and $b$, and $t^{\scriptscriptstyle 0}_i$ is a hopping parameter of pristine MLP.
Let now consider the energy spectrum of strained MLP with the modified hopping parameters $t_1$ and $t_2$.
The new $k$-space Hamiltonian 
of the strained MLP is given by 
\begin{flalign}
 \mathcal H_{\rm strained}(\bm k)& = \begin{pmatrix} 0 & h_{12}(\bm k)+h_{14}(\bm k) \\
 h^*_{12}(\bm k)+h^*_{14}(\bm k)& 0 
\label{hii}
\end{pmatrix},
\end{flalign}
where $h_{12}$ and $h_{14}$ are now defined in terms of the modified hopping parameters.
Diagonalizing this Hamiltonian at the $\Gamma$ point gives the band gap as
\begin{small}
\begin{eqnarray}
E_g&=&(4t^{\scriptscriptstyle 0}_1+2t^{\scriptscriptstyle 0}_2)-\sum_j(8\alpha^1_jt^{\scriptscriptstyle 0}_1+4\alpha^2_jt^{\scriptscriptstyle 0}_2)\varepsilon,
\label{eg01}
\end{eqnarray}
\end{small}where $j$ denotes the summation over $x$, $y$ components. The first parentheses is the unstrained 
band gap i.e. $E^{\scriptscriptstyle 0}_g=1.85$~eV and the second one indicates the  structural dependent
values of changes in the band gap 
due to the applied strain. Using the numerical values of the structural parameters in
Eq.~(\ref{eg01}), the band gap evolution of MLP in the presence of compressive
equi-biaxial strain is a linear function as
\begin{eqnarray}
E_g=E^{\scriptscriptstyle 0}_g-s_{\scriptscriptstyle 0}\varepsilon,
\label{eg}
\end{eqnarray}
where $s_{\scriptscriptstyle 0}=-8.2$~eV.
This equation shows that by applying  
compressive equi-biaxial strain, the energy gap gradually decreases as well as the effective gap $\delta$.
Now, the continuum model~(\ref{Hk21}) is rewritten in terms of new coefficients
$\alpha_\varepsilon$, $\beta_\varepsilon$ and 
$\gamma_\varepsilon$, whose values are
\begin{eqnarray}
 \alpha_\varepsilon &=&\alpha_{\scriptscriptstyle 0}-\frac{1}{2}s_{\scriptscriptstyle 0}\varepsilon ,\notag \\
 \beta_\varepsilon &\approx&\beta_{\scriptscriptstyle 0}[1-2\varepsilon(s_{\scriptscriptstyle 1}-1)]\approx\beta_{\scriptscriptstyle 0}, \notag \\
 \gamma_\varepsilon&\approx&\gamma_{\scriptscriptstyle 0}+
 \varepsilon[-2(1-2s_{\scriptscriptstyle 1})t^{\scriptscriptstyle 0}_1r^{\scriptscriptstyle 0}_{1x}
 +(1-2s_{\scriptscriptstyle 2})t^{\scriptscriptstyle 0}_2r^{\scriptscriptstyle 0}_{2x}],
 \label{str11}
\end{eqnarray}
 \begin{figure}[!t]
\centering
{\includegraphics[angle=0,width=0.46\textwidth]{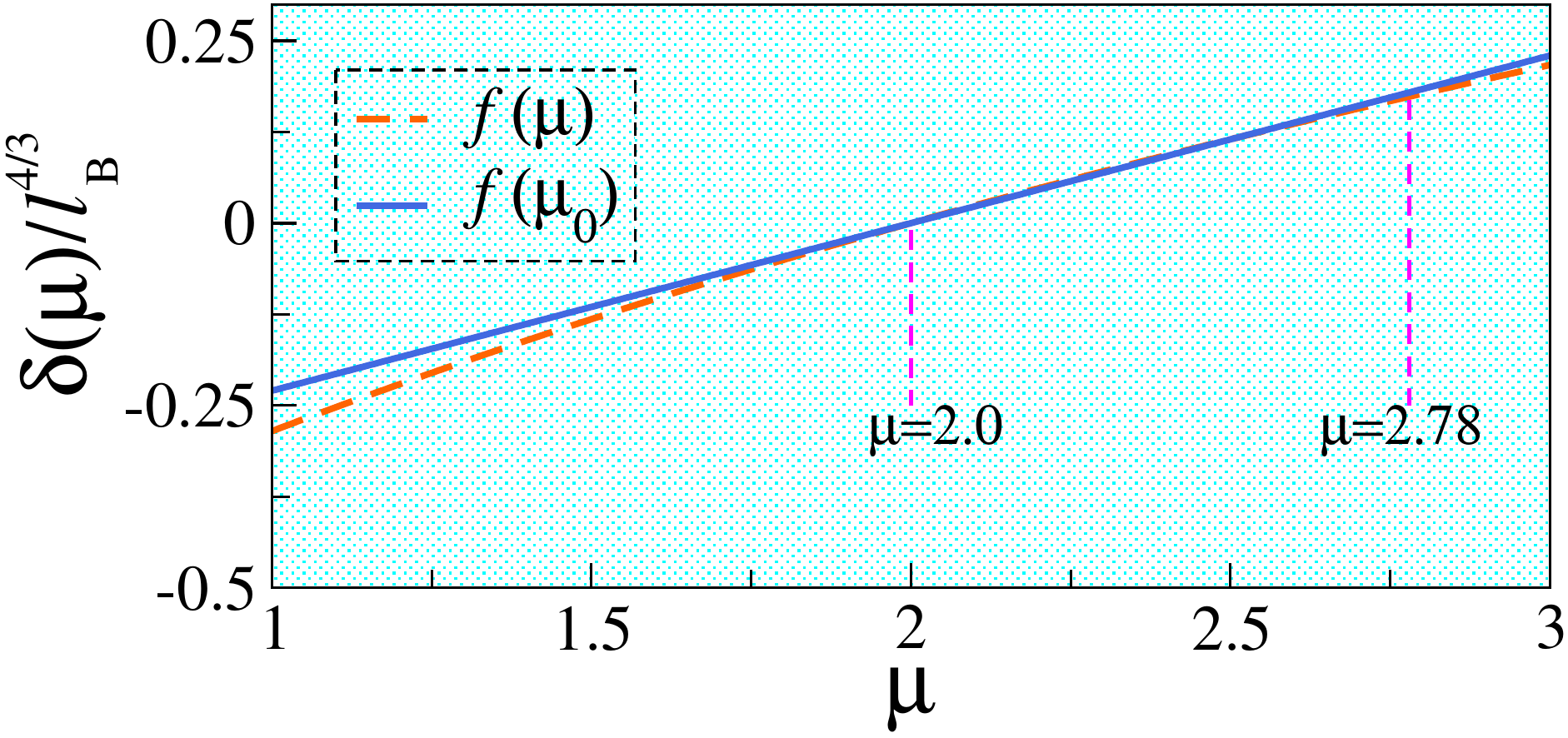}}
 \caption{Behavior of $\delta(\mu)$ function (in unit of $\ell_B^{4\texttt{\small/}3}$)
 and for the special case where $f(\mu)$ is replaced by $f(\mu_{\scriptscriptstyle0})$.}
\label{delta}
\end{figure}
\begin{figure}[!t]
\centering
{\includegraphics[angle=0,width=0.28\textwidth]{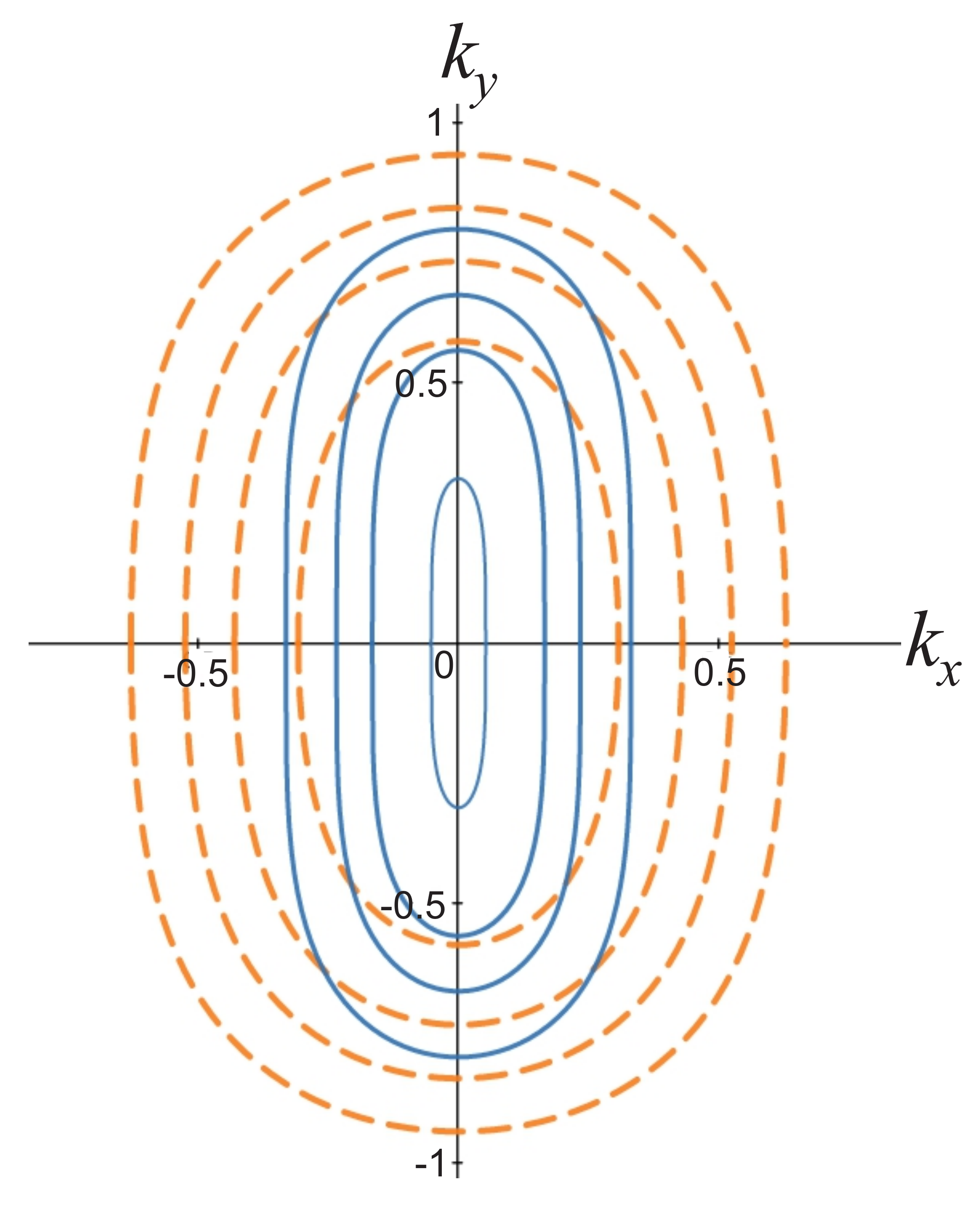}}
 \caption{Dashed orange and solid blue ovals depict some typical $k$-space cyclotron orbits for MLP at $\varepsilon=0$ and $\varepsilon=\varepsilon_c$, respectively.}
\label{oval}
\end{figure}where we have defined $s_{\scriptscriptstyle1}=\alpha_x^1+\alpha_y^1\approx1.0$, 
and $s_{\scriptscriptstyle2}=\alpha_x^2+\alpha_y^2\approx0.12$.
This leads to the strain dependent continuum spectrum of electrons and holes as 
\begin{equation}
 E(k_x,k_y,\varepsilon)=\pm\sqrt{(\alpha_\varepsilon+\beta_\varepsilon k_y^2)^2+\gamma_\varepsilon^2k_x^2},
\label{Ek}
\end{equation}
which is defined over the modified BZ. The energy spectra along the $\Gamma$-Y and $\Gamma$-X directions 
are given by 
\begin{equation}
 E(k_y,\varepsilon)=\pm(\alpha_\varepsilon+\beta_\varepsilon k_y^2),
\label{Ek11}
\end{equation} and 
\begin{equation}
 E(k_x,\varepsilon)=\pm\sqrt{\alpha_\varepsilon^2+\gamma_\varepsilon^2k_x^2},
\label{Ek12}
\end{equation}
respectively. Equation~(\ref{Ek11}) explicitly shows that for an arbitrary strength of
applied strain, the parabolic nature of bands along the $\Gamma$-Y direction  remains unchanged,
whereas, the Eq.~(\ref{Ek12}) implies that by increasing its strength, the relativistic nature of bands along the $\Gamma$-X direction becomes more evident. For the critical 
strain value $\varepsilon_c$,  the energy gap is closed and
charge carriers  become massless relativistic particles as  
\begin{equation}
 E(k_x,\varepsilon_c)=\pm \gamma_{\varepsilon_c}k_x.
\label{Ek13}
\end{equation}
Let us now consider the effect of applying a perpendicular magnetic field.
In terms of the new strain-modified hopping parameters one can rewrite  Eq.~(\ref{delta1}) as 
\begin{eqnarray}
\delta(\mu)&=&(\mu-2)f(\mu)^{-1}\ell_B^{4/3},\notag \\
f(\mu)&=&( r_{1y}r_{2x}\mu+2r_{1y}r_{1x})^{2/3},
\end{eqnarray}
 where $\mu=|t_2\texttt{\small/}t_1|$ shows the ratio of two strain modified hopping parameters. 
 For unstrained phosphorene we have $\mu_{\scriptscriptstyle 0}\approx2.78$. By applying strain
 up to the critical value $\varepsilon_c$, the values of $\mu$ ranges from $2.78$ to $2$. 
 As can be seen in Fig.~\ref{delta}, the behavior of
 function $\delta(\mu)/\ell_B^{4\texttt{\small/}3}$ in this range of $\mu$ is similar to the case
 in which $f(\mu)$ is 
 replaced by
 $f(\mu_{\scriptscriptstyle0})$. As a result, by increasing the strength of applied strain,
 the behavior of strain dependent potential $V_{l\Scale[0.6]{-}eff}(X)$
 is mainly determined by the magnitude of $\mu-2$ which is proportional to the gap of the system.
 This indicates that the effective gap parameter dramatically affects the ratio of the major to the minor axis ($a_{ov}\texttt{\small/}b_{ov}$) of the oval shape $k$-space cyclotron orbits in a GSD material. In other words, besides the curvature of energy bands (note that the Berry's phase zero) in a GSD system,  the magnitude of the band-gap parameter and its interplay with quartic plane cyclotron orbits determines the characteristics of LLs. 
 We consider the ratio $a_{ov}\texttt{\small/}b_{ov}$ as a criterion to represent the extent of anisotropy in cyclotron orbits. Figure~\ref{oval} displays some typical $k$-space cyclotron orbits for $\varepsilon=0$ (dashed orange orbits) and $\varepsilon=\varepsilon_c$ 
 (solid blue lines orbits).  From the numerical values of 
 $\alpha_\varepsilon$, $\beta_\varepsilon$ and  $\gamma_\varepsilon$ at different biaxial strains,  one can readily check that the parameter $\beta_\varepsilon$ does not change 
 the ratio $a_{ov}\texttt{\small/}b_{ov}$ and 
 the effect of $\gamma_\varepsilon$ is negligible compared to  
 $\alpha_\varepsilon$. Hence, the degree of anisotropy of $k$-space orbits is mainly decided by the band-gap  parameter. Therefore,  we would like to highlight the role of the effective gap and its interplay with the extent of anisotropy of oval shape cyclotron orbits 
in determining characteristic features of LLs for a GSD system.
The numerically obtained LLs of MLP as 
a function of equi-biaxial strain are shown in  Fig.~\ref{ll_mono}. Figures~\ref{ll_mono}(b) and (c) show the formation of equidistant LLs from low-energy band spectrum of unstrained phosphorene  and  non-equidistant LLs from that of strained phosphorene at the critical value $\varepsilon_c$,
respectively.
Indeed, the mentioned interplay manifests in the effective potential Eq.~(\ref{v_eff}), and 
at the critical value $\varepsilon_c$ it gives
\begin{equation}
V_{eff}(X)=X^4+2X\bm\sigma_z.
\label{v_eff}
\end{equation}
It is shown that the effect of the linear term  in this potential is 
negligible and the effective potential is in fact related to a 
quartic oscillator~\cite{dietl2008new}.  I this case, 
the quantized energy levels follows the field dependence of
$[(n+1\texttt{\small/}2)B]^{2\texttt{\small/}3}$~\cite{dietl2008new}. 
Therefore, by applying equi-biaxial strain and consequently the evolution of the band dispersion including both the band-gap parameter and the curvature of energy bands,  one can continuously evolve the field dependence of LLs from a linear to $B^{2\texttt{\small/}3}$ dependence.
In other words, these results emphasize the role of the band-gap parameter and its competitive nature with an anisotropic band spectrum in semi-Dirac systems that very likely emerge in many of their properties like magnetotransport, optical conductivity, dielectric function, and thermoelectric performance.
\section{Summary}
In summary, we have highlighted the role of the band-gap parameter and its 
interplay/rivalry with the bands curvature in characteristic features of LLs for a typical GSD system. In the case of pristine MLP, based on the continuum low-energy Hamiltonian and by deriving an effective one-dimensional Schrödinger equation, we demonstrated that the rivalry between the band anisotropy and the band-gap parameter always leads to the winning of the latter even for a very high-field regime. This resolves the seeming contradiction of the conventional linear field dependence of LLs in the presence of ellipse-like $k$-space cyclotron orbits, where one expects the anomalous $B^{2\texttt{\small/}3}$ field dependence.
Hence, our findings have convincingly established that the LLs in this GSD material linearly scales with the magnetic field, similar to a conventional 2D semiconductor electron gas.
We argued that the form of the effective potentials is sensitive to a dimensionless band-gap that is tunable by structural parameters. In addition, we showed that for this GSD system, biaxial compressive strains preserve the parabolic nature of the bands along the $\Gamma$-Y direction, while strengthen the relativistic nature of the bands along the $\Gamma$-X direction. At a critical strain value $\varepsilon_c$, the energy gap is closed and electrons and holes become massless relativistic particles. 
Therefore, by gradual suppression of the band-gap contribution in the mentioned interplay, one can continuously evolve the field dependence of LLs in a GSD system to that of a ZGSD one.
In addition to the importance of anisotropy in the low-energy bands of a GSD sample, our results have given particular emphasis on the role of the band-gap parameter that is very likely reflected in many of the physical properties of the system.

\bibliography{references}

\end{document}